%% file: main.tex
\documentclass[10pt,conference]{IEEEtran}
\IEEEoverridecommandlockouts
\usepackage{enumitem}
\usepackage{cite}
\usepackage{amsmath,amssymb,amsfonts}
\usepackage{algorithmic}
\usepackage{graphicx}
\usepackage{textcomp}
\usepackage{xcolor}
\usepackage{hyperref}
\usepackage{subcaption}
\usepackage{caption}
\def\BibTeX{{\rm B\kern-.05em{\sc i\kern-.025em b}\kern-.08em
    T\kern-.1667em\lower.7ex\hbox{E}\kern-.125emX}}
\begin{document}

\title{Towards Detecting Prompt Knowledge Gaps for Improved LLM-guided Issue Resolution}


\author{
\IEEEauthorblockN{Ramtin Ehsani}
\IEEEauthorblockA{Drexel University \\
Philadelphia, PA, USA \\
ramtin.ehsani@drexel.edu}
\and
\IEEEauthorblockN{Sakshi Pathak}
\IEEEauthorblockA{Drexel University \\
Philadelphia, PA, USA \\
sp3856@drexel.edu}
\and
\IEEEauthorblockN{Preetha Chatterjee}
\IEEEauthorblockA{Drexel University \\
Philadelphia, PA, USA \\
preetha.chatterjee@drexel.edu}
}

\maketitle

\begin{abstract}
Large language models (LLMs) have become essential in software development, especially for issue resolution. However, despite their widespread use, significant challenges persist in the quality of LLM responses to issue resolution queries. LLM interactions often yield incorrect, incomplete, or ambiguous information, largely due to knowledge gaps in prompt design, which can lead to unproductive exchanges and reduced developer productivity. 

In this paper, we analyze 433 developer-ChatGPT conversations within GitHub issue threads to examine the impact of prompt knowledge gaps and conversation styles on issue resolution. We identify four main knowledge gaps in developer prompts: \textit{Missing Context, Missing Specifications, Multiple Context}, and \textit{Unclear Instructions}. Assuming that conversations within closed issues contributed to successful resolutions while those in open issues did not, we find that ineffective conversations contain knowledge gaps in 44.6\% of prompts, compared to only 12.6\% in effective ones. Additionally, we observe seven distinct conversational styles, with \textit{Directive Prompting}, \textit{Chain of Thought}, and \textit{Responsive Feedback} being the most prevalent. We find that knowledge gaps are present in all styles of conversations, with \textit{Missing Context} being the most repeated challenge developers face in issue-resolution conversations.

Based on our analysis, we identify key textual and code-related heuristics—\textit{Specificity}, \textit{Contextual Richness}, and \textit{Clarity}—that are associated with successful issue closure and help assess prompt quality. These heuristics lay the foundation for an automated tool that can dynamically flag unclear prompts and suggest structured improvements. To test feasibility, we developed a lightweight browser extension prototype for detecting prompt gaps, that can be easily adapted to other tools within developer workflows.  
\end{abstract}


\input{sections/01_intro}
\input{sections/02_dataset}

\input{sections/03_methodology}
\input{sections/04_results}
\input{sections/07_feasibility}
\input{sections/05_threats}
\input{sections/06_background}
\input{sections/08_conclusion}

\bibliographystyle{IEEEtran}
\bibliography{ref}

\end{document}

%% file: sections/01_intro.tex
\section{Introduction}
\label{sec:intro}
Large language models such as ChatGPT, Gemini, and Claude have become crucial tools for software development. The 2023 JetBrains survey, which gathered responses from 26k developers across 196 countries, 77\% (i.e., three out of four developers) use ChatGPT for daily tasks~\cite{jetbrains_survey}. 
LLMs are transforming the way developers approach problem-solving, particularly in issue resolution~\cite{hou2024large, wu2023large}. 
Developers seek guidance from these models to troubleshoot and refine solutions. By providing real-time feedback, helping to debug, and accelerating problem-solving, LLMs have become integral to the issue resolution process~\cite{conversational_prog_blog, Ross_2023}.

Despite their popularity, several concerns remain
regarding the quality of LLM responses to issue resolution-related queries. 
Recent studies have shown that these interactions often yield incomplete, ambiguous, or incorrect information~\cite{Li2023AlwaysNA, mondal2024enhancing, zhou2024exploringproblemscausessolutions, liu2024exploringevaluatinghallucinationsllmpowered}. LLM responses are highly sensitive to the information provided in the prompts~\cite{battle2024unreasonableeffectivenesseccentricautomatic, 10.1145/3640543.3645200}. Therefore, \textit{prompt knowledge gaps} (e.g., missing context, unclear instructions) 
play a critical role in shaping these interactions. These gaps can lead to irrelevant responses or even hallucinations~\cite{mondal2024enhancing}. Knowledge gaps in prompt design also lead to multiple back-and-forth interactions, resulting in unsuccessful conversational outcomes and decreased developer productivity~\cite{hao2024empirical}.
These challenges highlight the need for a deeper understanding of how prompt gaps and styles affect LLM-driven issue resolution.

Previous research has examined gaps in developer-LLM conversations across different software engineering tasks, identifying factors that prolong conversations and require multiple prompts for developers to obtain helpful responses~\cite{10.1145/3640543.3645200, ma2024saywantteaching, kim2023betterexploringpromptingstrategy}. However, these studies do not identify the unique challenges and knowledge gaps specific to issue resolution, such as the need for precise problem descriptions and detailed error messages. 
A recent study analyzing 85 ChatGPT conversations related to issue resolution reveals 11 types of gaps in both developer prompts and ChatGPT's responses that contribute to longer exchanges~\cite{mondal2024enhancing}. 
Despite these findings, existing research has yet to systematically examine how knowledge gaps in developer prompts impact the effectiveness of LLM-driven issue resolution. 
While prompt engineering methods aim to refine LLM responses by optimizing phrasing, structure, and style~\cite{ma2024saywantteaching}, 
they often fail to address a deeper challenge: providing actionable, targeted guidance that empowers developers to create more effective prompts. As a result, effective issue resolution still heavily depends on developers’ ability to identify the knowledge gap, and incorporate the necessary information in the prompts.

In this paper, we analyze 433 developer-ChatGPT conversations shared within GitHub issue threads to investigate the impact of \textit{prompt knowledge gaps} and \textit{conversation styles} on issue resolution. 
We focus on knowledge gaps, such as identifying missing or ambiguous content, that impact the effectiveness of developer-LLM discussions related to issue resolution.
By identifying key textual and code-related heuristics that are related to these gaps, we explore patterns in conversations that are associated with successful issue closure on GitHub. 
Our objective is to identify actionable heuristics that can inform the design of a tool to dynamically flag unclear or incomplete prompts, and suggest improvements in the form of structured templates. 
Toward that goal, we investigate the following research questions:

\noindent
\textbf{RQ1:} \textit{How do prompt knowledge gaps and conversation styles influence the progression and effectiveness of developer-ChatGPT conversations in issue resolution?}
We annotate each developer-ChatGPT conversation with four categories of knowledge gaps and seven styles of conversations. By analyzing the content and discourse of each conversation, we investigate their influence on the GitHub issue status (open vs. closed). We find that developers use different styles and techniques such as \textit{Chain of Thought} and \textit{Directive Prompting} to interact with ChatGPT, however, knowledge gaps persist across all styles. Developers struggle with providing the right context, with \textit{Missing Context} emerging as the most common issue associated with unsuccessful conversations.

\noindent
\textbf{RQ2:} \textit{What heuristics can be used to automatically measure the prompt knowledge gaps?}
Based on the results of RQ1, we design three categories of textual and code-related heuristics (\textit{Contextual Richness, Specificity, and Clarity}) that capture the nuances of knowledge gaps in prompts when using ChatGPT for issue resolution. We found that providing short code snippets, additional information such as links to documentation, and error messages in the prompt, while maintaining the conversation on the same topic can lead to more effective issue resolution conversations. These heuristics provide a foundation for designing tools for automatic detection of prompt gaps.  

To demonstrate the feasibility of using our RQ2 heuristics in an automated tool, we develop a lightweight prototype. Implemented as a browser extension, this prototype can be adapted to other tools within developer workflows. Our work takes the first step towards automated prompt knowledge gap detection in LLM-aided issue-resolution conversations. By providing targeted, actionable suggestions, this tool could help developers proactively enhance prompt quality.
The key contributions in this paper are summarized as follows:
\begin{itemize}[leftmargin=*]
    \item We present a manually annotated dataset of developer-ChatGPT conversations focused on issue resolution, with annotations for prompt styles and knowledge gaps as the conversation progresses.
    \item We conduct a comprehensive analysis of these conversations, identifying common knowledge gaps and styles, and uncovering key heuristics that are strongly associated with the effectiveness of the interactions towards issue resolution.
    \item 
    We develop the first prototype for automatically detecting knowledge gaps in prompts during ChatGPT-based issue resolution, offering tailored templates to improve prompt quality and enhance the likelihood of successful outcomes.
\end{itemize}

%% file: sections/02_dataset.tex
\section{Dataset}
\label{sec:dataset}

We use the DevGPT dataset~\cite{xiao_devgpt_2024} for our analysis. This dataset contains developer-ChatGPT conversations that are publicly shared through links on platforms such as GitHub and Hackernews. These conversations are generated using OpenAI's web-browser platform of ChatGPT, which utilizes either GPT-3.5 or GPT-4. Since the focus of our study is issue resolution, we selected conversations shared within GitHub issue threads. 
These conversations contain a wide variety of queries directed at ChatGPT, including how-to questions, advanced programming guidance, inquiries about frameworks, and high-level design recommendations.~\cite{MohamedChatting, SagdicDiscussion, hao2024empirical}.
The conversations in this dataset often serve as references for potential solutions or helpful context~\cite{hao2024empirical}, and they are inherently related to the issues because developers intentionally share them as resources they believe might assist in resolving the problem.
We further analyzed the dataset and observed that the queries and subtasks presented to ChatGPT vary from straightforward ones such as API usage and syntax fixes to more complex debugging and multi-threading issues. Simpler tasks required fewer interactions and were resolved more effectively, while complex ones were more challenging. For example, resolving syntax errors like ``How do I fix a syntax error in Python?" was far more straightforward than diagnosing a segmentation fault in C. While task difficulty impacts ChatGPT’s performance, our focus remains on assessing how providing sufficient detail within prompts influences issue resolution regardless of the difficulty of issues. In addition, the diversity of subtasks covered in these conversations allows us to generalize our findings to a wide range of issue-resolution challenges.

Each dataset entry contains the ChatGPT link, the associated GitHub issue, the full conversation comprising each prompt and its corresponding ChatGPT response, and the saved HTML content of the conversation. The original dataset consists of 636 entries. 
We filtered out duplicate entries and non-English conversations using Python’s \textit{lingua-py} library~\cite{stahl_pemistahllingua-py_2024}. 
Code snippets and error messages were replaced with [CODE] and [ERROR] tags, respectively, and separated from the text. ChatGPT responses structure code snippets in quote blocks, allowing for easy replacement using RegEx, while developer prompts often do not. Following previous studies, to detect unstructured code and error messages in prompts, we used \textit{GPT-4}~\cite{Oedingen_2024}. One of the authors manually validated the dataset to ensure accuracy. Our final dataset comprises 433 developer-ChatGPT conversations shared within 400 unique GitHub issues.

%% file: sections/03_methodology.tex
\section{Methodology}
\label{sec:method}

\subsection{RQ1: How do prompt knowledge gaps and conversation styles influence the progression and effectiveness of developer-ChatGPT conversations in issue resolution?}

We annotate 433 developer-ChatGPT conversations, focusing on two main aspects: prompt knowledge gaps (i.e., deficiencies in the content of prompts) and conversation styles (i.e., the techniques developers used to communicate with ChatGPT). By annotating the dataset according to these two aspects, we assess how knowledge gaps and conversation styles contribute to the effectiveness of issue resolution. 
We assume that conversations within closed issues likely contributed to successful resolutions, while those within open issues did not effectively aid in resolving the issues. This approach is the best available option for evaluating the relationship between prompt knowledge and issue resolution. Our study focuses on understanding how prompt quality is associated with the likelihood of issue resolution. Given this, the status of the issue (open or closed) provides a direct and practical measure of effectiveness.

We followed a qualitative content analysis approach~\cite{AzungahQualitative}, combining both deductive and inductive coding methods. We began with a set of predefined categories for prompt knowledge gaps and conversation styles, derived from existing taxonomies and literature. Using this strategy, two authors of this paper independently annotated an initial subset of conversations to capture prompt gaps and conversation styles. Through iterative coding and discussion, we refined the categories, leading to the creation of modified taxonomies for both prompt knowledge gaps and conversation styles for issue resolution. This inductive refinement allowed us to adapt our categories based on observed data patterns, enhancing the validity of our coding scheme. To ensure reliability, we calculated Cohen’s Kappa scores at each stage, achieving strong inter-rater reliability in the final round, and further validated our consistency through a blinded sample check. Next, we discuss the evolution of the taxonomy and further details on the annotation procedure. 

To identify the prompt knowledge gaps, we started with Mondal et al.'s prompt gap taxonomy consisting of a total of 11 categories: \textit{Missing Specifications, Different Use Cases, Incremental Problem Solving, Exploring Alternative Approaches, Wordy Response, Additional Functionality, Erroneous Response, Missing Context, Clarity of Generated Response, Inaccurate/Untrustworthy Response, and Miscellaneous}~\cite{mondal2024enhancing}. 
To identify the conversation styles, we started with 18 categories: \textit{Meta Language Creation, Output Automator, Persona, Visualization Generator, Template, Skeleton of Thought, Chain of Thought, Tree of Thought, Fact Check List, Meta-prompting, Reflection, Responsive Feedback, Question Refinement, Alternative Approaches, Cognitive Verifier, Refusal Breaker, Game Play}, and \textit{Few-shot Learning}, drawn from the literature on interaction styles with LLMs~\cite{white2023promptpatterncatalogenhance, fagbohun2024empiricalcategorizationpromptingtechniques, noauthor_prompt_2024}.
Two authors independently reviewed an initial set of 50 ChatGPT conversations. We annotated the first prompt in each conversation to capture its initial gaps, while subsequent prompts were annotated based on new information the developers provided, allowing us to observe how knowledge gaps evolved through the conversation. After identifying prompt knowledge gaps, each conversation was categorized with a conversation style that reflected the overall interaction across all prompts. The inter-rater agreement for the first round of annotations was Cohen's Kappa of 0.62 for gaps and 0.48 for styles, indicating moderate agreement~\cite{keppaArticle}.
In the second round of annotation, the authors revisited the initial 50 conversations along with 50 additional ones. The final Cohen's Kappa agreement was 0.84 for gaps and 0.72 for styles, both reflecting strong inter-rater reliability~\cite{keppaArticle}. 
Conflicts in the annotations were iteratively discussed and resolved collaboratively with both annotators contributing equally, leading to a refinement of the initial categories. Since, we had high inter-rater agreement after the second iteration, the rest of the dataset was split among the two researchers to complete the annotation independently.

Based on our iterative discussions, for conversation styles seven categories were discarded because they were not present in our dataset. These categories were \textit{Meta Language Creation, Output Automator, Visualization Generator, Fact Check List, Cognitive Verifier, Refusal Breaker,} and \textit{Game Play}.
In addition, eight categories were merged into three categories because they represented the same style: \textit{Responsive Feedback, Meta-prompting, Reflection, Question Refinement} into \textit{Responsive Feedback}; \textit{Template} and \textit{Skeleton of Thought} into \textit{Template}; and \textit{Tree of Thought} and  \textit{Alternative Approaches} into \textit{Tree of Thought}. The final set included six categories, plus one additional style (Directive Prompting) derived from our coding.
The prompt knowledge gaps were consolidated into three main categories, with an additional gap (Unclear Instructions) emerging from our open coding process~\cite{corbin_basics_2008}. The categories discarded for prompt knowledge gaps were \textit{Different Use Cases, Incremental Problem Solving, Exploring Alternative Approaches, Wordy Response, Additional Functionality, Erroneous Response, Clarity of Generated Response,} and \textit{Inaccurate/Untrustworthy Response}.

Additionally, to ensure our annotations were not biased based on conversations' status (open vs. closed), we sampled 50 conversations after the annotation, hiding their status and redoing the annotation to see if we identified different gaps or styles in the conversations. In only 4 conversations (1 open and 3 closed) we identified additional gaps in prompts indicating that our annotations were consistent. 

We now present the refined taxonomies. 
\underline{Prompt Knowledge Gap Categories:}
As summarized in Table \ref{tab:prompt_gaps}, we categorize prompt knowledge gaps into four groups: \textit{Missing Context}, \textit{Missing Specification}, \textit{Unclear Instruction}, and \textit{Multiple Context}.
These gaps are essential for evaluating whether the developer provided enough information and clarity for ChatGPT to understand and resolve the issue.

\begin{table}[h!]
\centering
\caption{Prompt Knowledge Gaps in Developer-ChatGPT Conversations}
\label{tab:prompt_gaps}
\resizebox{\columnwidth}{!}{%
\begin{tabular}{|l|p{5cm}|}
\hline
\textbf{Category}            & \textbf{Description}                                               \\ \hline                                                         
Missing Context     & Lacks essential details, such as goals, previous attempts, or project info.                                                      \\ \hline
Multiple Context   & Introduces multiple issues without clear separation, leading to confusion.
\\ \hline
Unclear Instructions & Instructions are vague or open to multiple interpretations, leading to ineffective responses.                                   \\ \hline
Missing Specification & Lacks critical technical information (e.g., programming language).                                                    \\ \hline
\end{tabular}
}
\end{table}

\noindent
\textbf{Context} refers to the background information that developers provide to help ChatGPT understand the problem. We identified gaps in context by looking for prompts that lacked sufficient background information. 
A prompt was labeled as \textit{Missing Context} if it did not provide essential details like the user’s end goals, prior attempts to solve the issue, codes and error logs, or relevant project information~\cite{ouyang2022traininglanguagemodelsfollow}. On the other hand, a prompt was classified as \textit{Multiple Context} when it introduced more than one distinct issue in the same conversation thread. This often leads to confusion in responses, as ChatGPT struggles to focus on one problem.

\noindent
\textbf{Instructions} refer to the explicit steps or actions that developers want ChatGPT to perform. Unambiguous instructions are crucial for obtaining relevant and accurate responses. We analyzed prompts to identify instances where the instructions provided were unclear or open to multiple interpretations. Unclear instructions with grammatical issues, misspellings, or anything that hinders the understanding of the instruction is classified as \textit{Unclear Instructions}. One such example is:\textit{``noe to how to run all togathor and display in website"}.

\noindent
\textbf{Specification} relates to the technical details and system requirements specific to the issue. Effective prompts should contain enough technical information, such as exact programming language, performance constraints, or versions of the frameworks to guide ChatGPT in generating precise solutions. We categorized prompts as having a \textit{Missing Specification} gap if they lacked essential technical details necessary for providing a meaningful solution.

\underline{Conversation Style Categories:} 
We identified seven conversation styles as follows: \textit{Persona}, \textit{Template}, \textit{Chain of Thought}, \textit{Tree of Thought}, \textit{Responsive Feedback}, \textit{Few-shot Learning}, and \textit{Directive Prompting}.

\noindent
\textbf{Persona} is a style where developers instruct ChatGPT to assume a specific role or perspective. By asking ChatGPT to “act as a cybersecurity expert” or “explain this as a mentor would,” developers can tailor responses to align with their specific needs. This style is particularly useful when the developer requires expert-level advice or wants the response framed in a particular way~\cite{white2023promptpatterncatalogenhance}.

\noindent
\textbf{Template} is a style where developers provide a predefined structure for ChatGPT to follow in its output. Developers often use this style when they need the response to adhere to a specific format, such as a documentation template or structured report. By giving ChatGPT a template to follow, developers ensure consistency in responses, particularly when the output must follow a standardized format~\cite{white2023promptpatterncatalogenhance}.

\noindent
\textbf{Chain of Thought} breaks down complex tasks into logical, sequential steps. Instead of asking ChatGPT for an immediate solution, the developer prompts the model to think through the problem step by step. This style is especially effective for multifaceted problems where each step needs to be carefully considered~\cite{ fagbohun2024empiricalcategorizationpromptingtechniques, noauthor_prompt_2024}.

\noindent
\textbf{Tree of Thought} expands on the Chain of Thought approach by encouraging ChatGPT to explore multiple possible solutions or pathways. In situations where there is more than one potential solution, this style allows developers to prompt ChatGPT to branch out and explore various scenarios or alternative strategies~\cite{ fagbohun2024empiricalcategorizationpromptingtechniques, noauthor_prompt_2024}.

\noindent
\textbf{Responsive Feedback} is a style where developers provide feedback directly within the prompting process to refine ChatGPT’s responses. For example, after receiving an initial output, the developer might give feedback such as “I like this part, but can you make it simpler?” This allows for iterative improvements and dynamic interaction, leading to more refined and tailored responses~\cite{white2023promptpatterncatalogenhance, fagbohun2024empiricalcategorizationpromptingtechniques}.

\noindent
\textbf{Few-shot Learning} is when developers provide a few examples within the prompt to illustrate their request. By including these examples, developers can help ChatGPT better understand the task and generate responses that align with their expectations~\cite{fagbohun2024empiricalcategorizationpromptingtechniques, noauthor_prompt_2024}.

\noindent
\textbf{Directive Prompting} is a style where we identified developers provide goals and the scope of issues to direct ChatGPT toward a specific outcome. This style is straightforward with developers knowing what they exactly want, which helps reduce ambiguity and ensures that the conversation stays focused on the desired solution.

\begin{figure*}[ht!]
    \centering
    \includegraphics[width=1\linewidth]{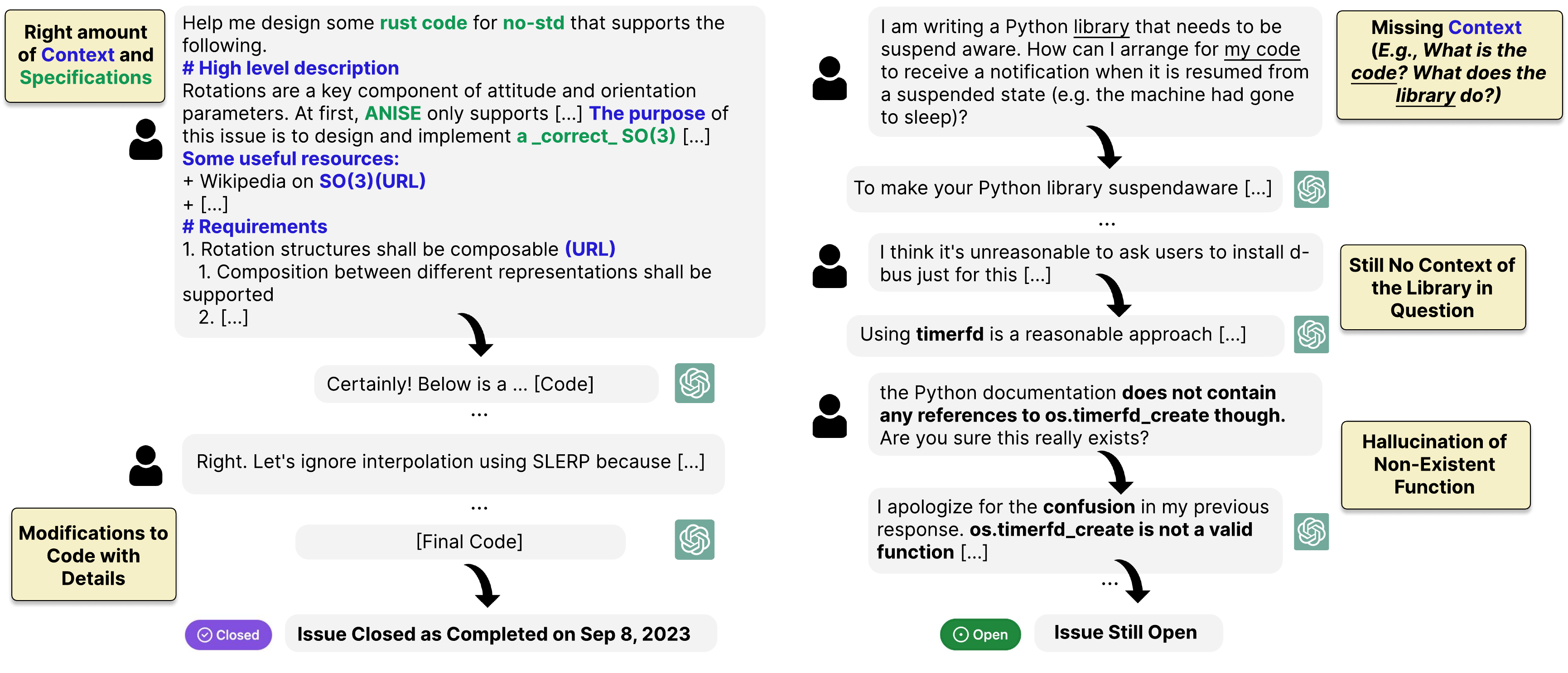}
    \caption{Example of Open vs. Closed Conversations: Closed conversation provides \textcolor{blue}{Context} and \textcolor{teal}{Specifications} to ChatGPT vs. the missing \textcolor{blue}{Context} in open conversation lead ChatGPT to hallucinate.}
    \label{fig:open_vs_closed}
    \vspace{-0.5cm}
\end{figure*}

\subsection{RQ2: What heuristics can be used to automatically measure the prompt knowledge gaps?}
Based on the results of RQ1 and further analysis of the content of developer prompts, we design three categories of heuristics: \textit{Specificity}, \textit{Contextual Richness}, and \textit{Clarity} (see Table \ref{tab:heuristics}). 
These heuristics are coming from our analysis of what constitutes knowledge gaps in developer prompts and were derived using NLP and code-related metrics to directly correspond to the knowledge gaps identified in prompts. \textit{Contextual Richness} helps identify \textit{Missing Context} and \textit{Multiple Context} by measuring the inclusion of code, external references, error messages, and other necessary information that developers may overlook. \textit{Specificity} addresses \textit{Missing Specification} by evaluating if the prompt contains enough technical details to frame the developers' requests. Finally, \textit{Clarity} tackles \textit{Unclear Instructions} by analyzing how cohesively instructions are structured, aiming to detect any ambiguous language that could lead to misunderstandings.

Additionally, to assess the impact of these heuristics on the effectiveness of issue resolution, we quantitatively evaluate them across conversations from open and closed issues using logistic regression models. This approach allows us to explore how these three heuristics are associated with successful outcomes, i.e., closed issues. 


\subsubsection{\textbf{Specificity}}
Specificity assesses the degree of detail in a developer’s prompts, focusing on how thoroughly technical requirements and specific requests are communicated. 
High specificity, such as indicating the programming language, library, or version, allows ChatGPT to better understand the problem and produce more accurate, relevant responses. 
We measure Specificity through three categories: \textit{Technical Keywords}, \textit{Conditional Phrasing}, \textit{Information Emphasis}.

\textit{Technical Keywords}: Inclusion of specific technical terms or commands can clarify the task and narrow the focus of the response. This could help ChatGPT align its output closely with the developer’s needs. 
We use two metrics: the frequency of software-specific terms, and named entities. 
We calculate the \textit{\#Software-specific Terms} using a pre-compiled list of morphological terms and software-related terms~\cite{7985684, dictionary_SE}. For \textit{\#Named Entities} we use Python's NLTK package~\cite{noauthor_nltk_nodate}.

\textit{Conditional Phrasing}: This metric evaluates how developers structure prompts to clarify task-specific requirements. We calculate the frequency of \#Constraints, \#Modifiers, and \#Subordinate Clauses using Python's NLTK. 
For instance, \textit{Subordinate Clauses} introduce additional details or conditions that refine the request (e.g., ``if the library is compatible with Python 3.8''). 

\textit{Information Emphasis}:
This metric captures how developers use repetition to maintain focus on key points. 
To capture this, we measure \#Repeated N-grams (n=2, 3) within the prompts to identify instances 
where developers reinforce important information.

\subsubsection{\textbf{Contextual Richness}}
We assess how much contextual and background information developers provide in their prompts, as this is crucial for ChatGPT to understand the issue at hand. Capturing contextual richness in the text is challenging since it varies widely depending on the problem~\cite{melo2020exploringcontextawareconversationalagents}. We categorize it into four key categories: \textit{Information Density}, \textit{Code Elements}, \textit{References}, and \textit{Verbosity}.

\textit{Information Density}: The more unique and concentrated the information, the richer the context~\cite{10.1145/3450503}. We measure this by calculating \#Unique Words in the prompt and the ratio of distinct words to the total word count (\#Unique Info).

\textit{Code Elements}: Concrete artifacts, such as code snippets and error messages, add substantial context. We measure this using the number of \#Code Snippets, \#Error Message, and Mean Size Code Snippets included in the prompts. To measure \#Code descriptions, we tokenize the code identifiers from the code snippets in each conversation and count the sentences mentioning these tokens~\cite{10.1145/3450503, 7962359}.

\textit{References}: Referring to APIs or URLs that are relevant to the issue could provide additional context and knowledge to understand the problem. We use regular expressions to count \#URLs within the prompt.

\textit{Verbosity}: Verbosity reflects the extent to which developers elaborate on the problem. Using Python’s package spaCy~\cite{noauthor_spacy_nodate}, we calculate the total number of prompts in a conversation, and measure the \#Words and \#Sentences. 

\subsubsection{\textbf{Clarity}}
Unclear prompts are those that are vague, ambiguous, or open to multiple interpretations. Prompts must be clear enough for ChatGPT to accurately interpret them. To assess the clarity of prompts, we use two categories: \textit{Readability} and \textit{Ambiguity}.

\textit{Readability}: We evaluate how easily the information can be read and understood. First, using Python's pyspellchecker~\cite{barrus_pyspellchecker_nodate}, we count the number of misspelled words (\#Misspellings). Using Python's spaCy~\cite{noauthor_spacy_nodate}, we identify incomplete sentences lacking a subject or object as \#Incomplete Sentences~\cite{10.1145/3450503}. Additionally, we compute two widely recognized readability metrics: the Flesch Reading Ease Score~\cite{flesch_nodate} and the SMOG Grade~\cite{scott_smog_2023}. The Flesch score gauges how easy a text is to read, with higher scores indicating simpler content. The SMOG Grade estimates the years of education needed to understand a text. We calculate both using Python's py-readability-metrics package.

\textit{Ambiguity}: To assess ambiguity, we analyze the \#Unresolved References in prompts, focusing on instances of unclear pronoun usage. Using spaCy's NeuralCoref~\cite{noauthor_spacy_nodate}, we identify cases where pronouns lack a clear antecedent. 
Additionally, we use Natural Language Inference (NLI) to capture deeper contextual confusion. Using the RoBERTa-MNLI model~\cite{liu2019roberta}, we measure how confidently it classifies relationships between sentences. This model generates an Entailment score (higher means less ambiguous), indicating if the text logically follows the context.

\begin{table*}[h]
\caption{Heuristics to Capture Knowledge Gaps in Prompts and their Range in Conversations of Open and Closed Issues}
\label{tab:heuristics}
\resizebox{\textwidth}{!}{%
\begin{tabular}{|c|l|l|l|c|c|}
\hline
\multicolumn{1}{|l|}{{\textbf{\begin{tabular}[c]{@{}l@{}}Knowledge \\ Gaps\end{tabular}}}} & {\textbf{Heuristic}} & {\textbf{Heuristic Categories}} & {\textbf{Metrics}} & \textbf{Open Issues} & \textbf{Closed Issues} \\ \cline{5-6} 
\multicolumn{1}{|l|}{} &  &  &  & \textbf{Range (min\textless{}median\textless{}max)} & \textbf{Range (min\textless{}median\textless{}max)} \\ \hline
{\begin{tabular}[c]{@{}c@{}}Missing\\ Specification\end{tabular}} & {Specificity} & Technical Keywords & \begin{tabular}[c]{@{}l@{}}\#Software-specific Terms\\ \#Named Entities\end{tabular} & \begin{tabular}[c]{@{}c@{}}0\textless{}8\textless{}289\\ 0\textless{}2\textless{}635\end{tabular} & \begin{tabular}[c]{@{}c@{}}0\textless{}8\textless{}185\\ 0\textless{}2\textless{}91\end{tabular} \\ \cline{3-6} 
 &  & Conditional Phrasing & \begin{tabular}[c]{@{}l@{}}\#Constraints\\ \#Modifiers\\ \#Subordinate Clauses\end{tabular} & \begin{tabular}[c]{@{}c@{}}0\textless{}0\textless{}13\\ 0\textless{}6\textless{}693\\ 0\textless{}0\textless{}31\end{tabular} & \begin{tabular}[c]{@{}c@{}}0\textless{}0\textless{}26\\ 0\textless{}6\textless{}351\\ 0\textless{}0\textless{}49\end{tabular} \\ \cline{3-6} 
 &  & Information Emphasis & \begin{tabular}[c]{@{}l@{}}\#Repeated 2-grams\\ \#Repeated 3-grams\end{tabular} & \begin{tabular}[c]{@{}c@{}}0\textless{}2\textless{}530\\ 0\textless{}0\textless{}307\end{tabular} & \begin{tabular}[c]{@{}c@{}}0\textless{}1\textless{}270\\ 0\textless{}0\textless{}176\end{tabular} \\ \hline
{\begin{tabular}[c]{@{}c@{}}Missing\\ Context\end{tabular}} & \multicolumn{1}{c|}{{Contextual Richness}} & Code Elements & \begin{tabular}[c]{@{}l@{}}\#Code Snippets\\ Mean Size Code Snippets\\ \#Error Message\\ \#Code Descriptions\end{tabular} & \begin{tabular}[c]{@{}c@{}}0\textless{}0\textless{}108\\ 0\textless{}0\textless{}9257\\ 0\textless{}0\textless{}47\\ 0\textless{}0\textless{}669\end{tabular} & \begin{tabular}[c]{@{}c@{}}0\textless{}0\textless{}77\\ 0\textless{}0\textless{}3493\\ 0\textless{}0\textless{}34\\ 0\textless{}0\textless{}94\end{tabular} \\ \cline{3-6} 
 & \multicolumn{1}{c|}{} & Information Density & \begin{tabular}[c]{@{}l@{}}First Prompt Length\\ \#Unique Info\\ \#Unique Words\end{tabular} & \begin{tabular}[c]{@{}c@{}}1\textless{}41\textless{}2297\\ 2\textless{}11\textless{}56\\ 3\textless{}39\textless{}1247\end{tabular} & \begin{tabular}[c]{@{}c@{}}5\textless{}40\textless{}1727\\ 1\textless{}11\textless{}96\\ 1\textless{}38\textless{}598\end{tabular} \\ \cline{1-1} \cline{3-6} 
{\begin{tabular}[c]{@{}c@{}}Multiple\\ Context\end{tabular}} & \multicolumn{1}{c|}{} & References & \#URLs & 0\textless{}0\textless{}16 & 0\textless{}0\textless{}10 \\ \cline{3-6} 
 & \multicolumn{1}{c|}{} & Verbosity & \begin{tabular}[c]{@{}l@{}}\#Words \\ \#Sentences\\ \#Total Prompt Count\end{tabular} & \begin{tabular}[c]{@{}c@{}}4\textless{}54\textless{}4837\\ 1\textless{}3\textless{}299\\ 1\textless{}2\textless{}42\end{tabular} & \begin{tabular}[c]{@{}c@{}}1\textless{}53\textless{}2490\\ 1\textless{}3\textless{}146\\ 1\textless{}2\textless{}30\end{tabular} \\ \hline
{\begin{tabular}[c]{@{}c@{}}Unclear\\ Instructions\end{tabular}} & {Clarity} & Readability & \begin{tabular}[c]{@{}l@{}}\#Misspellings\\ \#Incomplete Sentences\\ Flesch Reading Ease Score\\ SMOG Grade\end{tabular} & \begin{tabular}[c]{@{}c@{}}0\textless{}1\textless{}92\\ 0\textless{}0\textless{}97\\ -44.2\textless{}66.7\textless{}102.6\\ 0\textless{}7\textless{}17\end{tabular} & \begin{tabular}[c]{@{}c@{}}0\textless{}1\textless{}14\\ 0\textless{}0\textless{}50\\ -155.5\textless{}68.3\textless{}117.1\\ 0\textless{}7\textless{}19\end{tabular} \\ \cline{3-6} 
 &  & Ambiguity & \begin{tabular}[c]{@{}l@{}}\#Unresolved Reference\\ Entailment\end{tabular} & \begin{tabular}[c]{@{}c@{}}0\textless{}3\textless{}138\\ 0.001\textless{}0.10\textless{}0.98\end{tabular} & \begin{tabular}[c]{@{}c@{}}0\textless{}3\textless{}199\\ 0.002\textless{}0.10\textless{}0.97\end{tabular} \\ \hline
\end{tabular}%
}
\vspace{-0.5cm}
\end{table*}

%% file: sections/04_results.tex
\section{Results and Discussion}
\label{sec:results}
\subsection{RQ1: How do prompt knowledge gaps and conversation styles influence the progression and effectiveness of developer-ChatGPT conversations in issue resolution?}

Out of 433 conversations, 262 were linked within closed issues while 171 were related to open issues. In total, open issues had 749 prompts, while closed issues had 849 prompts. Although there are more conversations in closed issues, open issues had a higher average number of prompts per conversation. This suggests that conversations in open issues tend to take longer to reach a solution. 

\begin{figure}[h]
    \centering
    \includegraphics[width=0.7\linewidth]{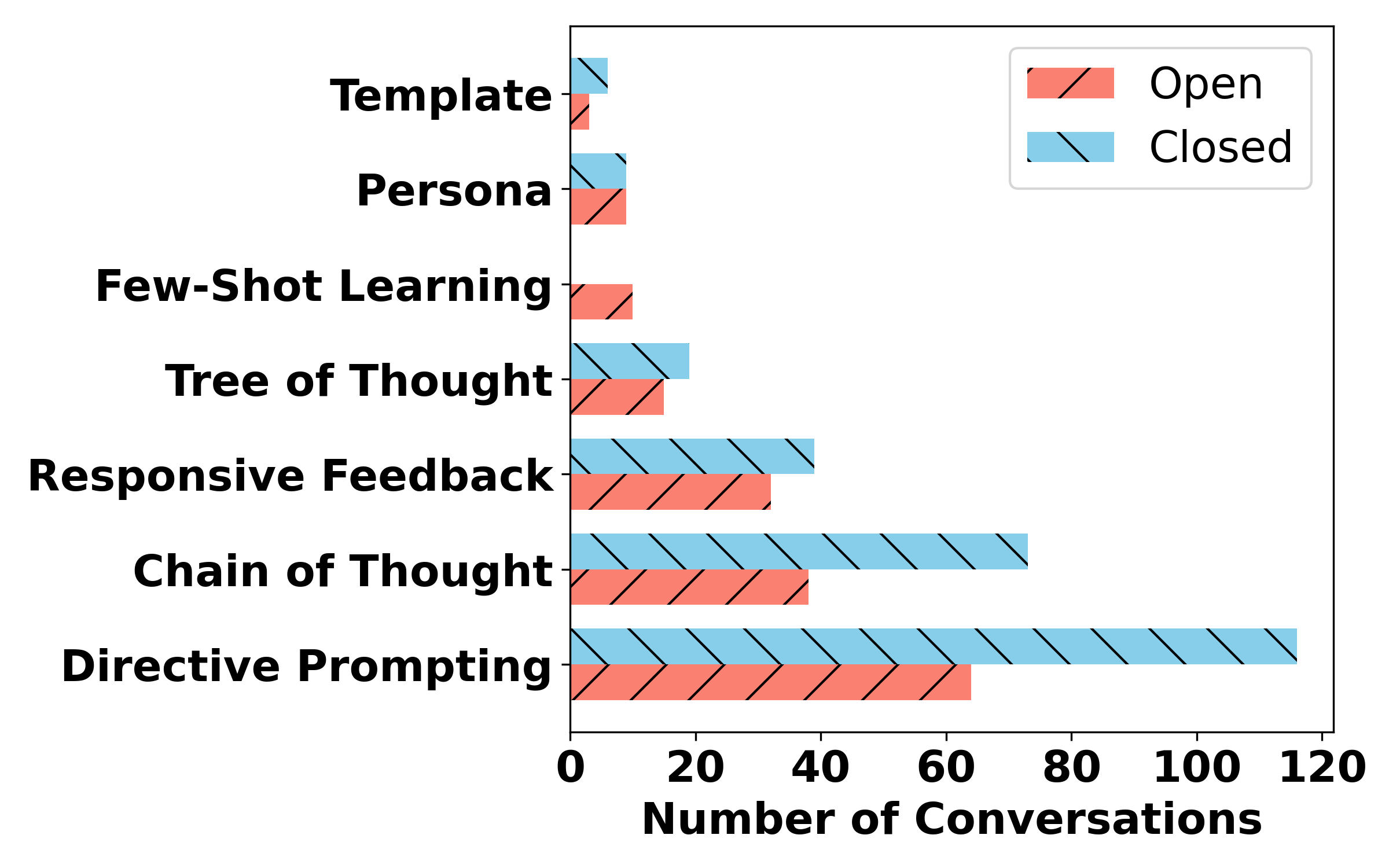}
    \caption{Styles of Conversations in Closed Vs. Open Issues}
    \label{fig:styles_open_vs_closed}
    \vspace{-0.3cm}
\end{figure}

Figure \ref{fig:styles_open_vs_closed} shows the frequency of the seven conversation styles across open and closed issues. 
The predominant styles of conversation in both open and closed issue threads were \textit{Directive Prompting, Chain of Thought, and Responsive Feedback}. Given the large sample size, we applied the independent t-test~\cite{kim_t_2015} (p-value\textless 0.05), which is robust to minor deviations from normality. The t-test showed no significant difference in the styles employed across open and closed issues (p-value=0.48), indicating that developers maintain a consistent approach when framing their questions for issue resolution. We also conducted the Shapiro-Wilk test~\cite{ghasemi_normality_2012} to confirm that the data follows a normal distribution. Additionally, the Mann-Whitney U test~\cite{Mann-Whitney} (p-value\textless 0.05) showed no significant differences either (p-value=0.84). \textit{Few-shot Learning} style of conversation was only noticed in open issues. This approach involved providing examples for ChatGPT to learn and generate relevant responses.  For example, in a conversation about adding JSP support programmatically to the code, the developer provided an example of what a code with this support might look like. 

We found a significant difference in the number of prompts with knowledge gaps: 334 in open issues compared to only 107 in closed ones. In open issues, the most common gap is \textit{Missing Context} (n=262), followed by \textit{Unclear Instructions} (n=66), \textit{Multiple Context} (n=45), and \textit{Missing Specification }(n=37). In closed issues, 742 of 849 prompts showed no gaps, but \textit{Missing Context} (n=77) remained the most frequent gap.

Providing the right context for an issue is the biggest challenge developers face when interacting with ChatGPT. We observed numerous cases where ChatGPT struggled to grasp the necessary context due to \textit{Missing Context}.
For example, Figure \ref{fig:open_vs_closed} shows a conversation within an open issue where a developer asked ChatGPT how to make their Python library suspend-aware.
However, they did not provide enough details about the library's functionality, the framework it was built on, and other critical information that was necessary to generate the correct answer. This resulted in ChatGPT hallucinating and using non-existent functions to compensate for the lack of information, leading to unhelpful answers to the developer's questions. Additionally, we observe that \textit{Missing context} is a critical issue in all styles of open-issue conversations, showing that no matter what style developers use, the problem of providing the right context still persists.

\begin{figure}[h]
    \centering
    \includegraphics[width=0.55\linewidth]{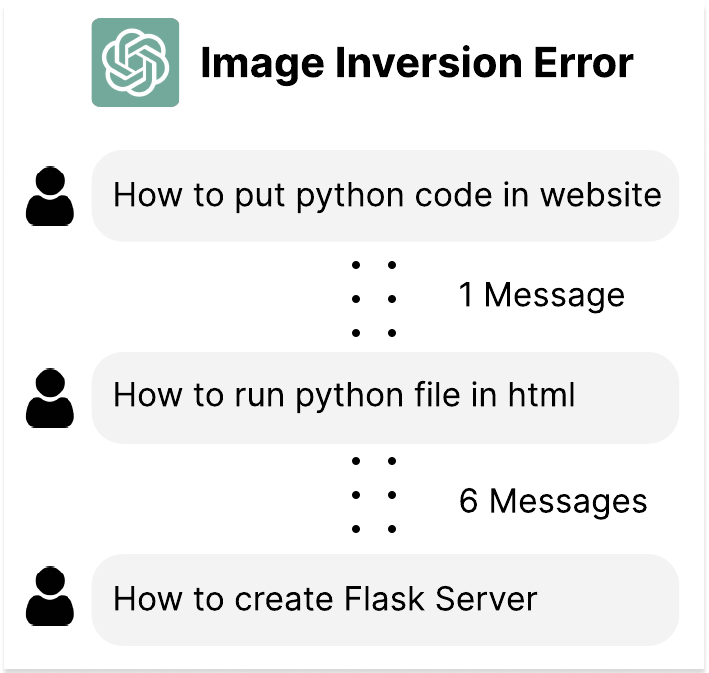}
    \caption{Multiple Context in a Conversation Linked to an Open Issue}
    \label{fig:mult}
    \vspace{-0.2cm}
\end{figure}

Table \ref{tab:gaps_prompts} shows the number and percentage of gaps per conversation style for both open and closed issues. Since closed issues have a higher total prompt count, the higher number of knowledge gaps within each category for open issues result in higher percentages. Almost all conversation styles in open issues exhibit at least one type of gap, with \textit{Chain of Thought} showing the highest number of gaps across all identified categories. Conversations adopting Chain of Thought prompting in closed conversations do not exhibit the gap of \textit{Multiple Context} (n=0), indicating that developers focus each discussion with ChatGPT on a single topic. However, in open issues, conversations with Chain of Thought prompting frequently suffer from \textit{Multiple Context} (n=43). In these conversations, developers often discuss their problems step by step, but unexpectedly shift the topic to something unrelated. For instance, as shown in Figure \ref{fig:mult}, problems such as embedding Python code in a website, running Python within HTML, and creating a Flask server are presented in one conversation thread. Changes in topics within the same conversation thread create confusion and lead to unclear assistance from ChatGPT. 
 
The other two conversation styles that are most influenced by prompt knowledge gaps across open and closed issues are Directive Prompting and Responsive Feedback. 
Overall, in closed issues, Directive Prompting is the style most affected by knowledge gaps, as it often lacks the context and background information needed to resolve complex issues. 
One other important difference between open and closed issues is the use of Responsive Feedback. We see noticeably fewer prompts with missing context in closed issues for this style. When providing feedback to ChatGPT, it is important to assess what ChatGPT is struggling with the most, and incorporate more context and information on the problem to get better answers from ChatGPT. For instance, in a conversation linked to a closed issue, a developer was trying to get help with writing code for SQLite database in Python to merge rows from different tables. After the first attempt, the developer was not satisfied with the answer provided by ChatGPT, but received the required answer at the next prompt by providing the schema in more detail: \textit{``[...]This is the table scheme of favorites: CREATE TABLE favorites[...]"}. 

\begin{table}[]
\caption{Number and Percentage of Prompts with Knowledge Gaps Per Conversation Style in Open (OP.) and Closed (CL.) Issues}
\label{tab:gaps_prompts}
\resizebox{\columnwidth}{!}{%
\renewcommand{\arraystretch}{1.2} 
\large 
\begin{tabular}{|l|cc|cc|cc|cc|}
\hline
\textbf{\begin{tabular}[c]{@{}l@{}}Style of \\ Conversation\end{tabular}} & \multicolumn{2}{c|}{\textbf{\begin{tabular}[c]{@{}c@{}}Missing\\ Context\end{tabular}}} & \multicolumn{2}{c|}{\textbf{\begin{tabular}[c]{@{}c@{}}Missing\\ Specification\end{tabular}}} & \multicolumn{2}{c|}{\textbf{\begin{tabular}[c]{@{}c@{}}Unclear\\ Instructions\end{tabular}}} & \multicolumn{2}{c|}{\textbf{\begin{tabular}[c]{@{}c@{}}Multiple\\ Context\end{tabular}}} \\ \cline{2-9} 
 & \multicolumn{1}{c|}{\textbf{OP.}} & \textbf{CL.} & \multicolumn{1}{c|}{\textbf{OP.}} & \textbf{CL.} & \multicolumn{1}{c|}{\textbf{OP.}} & \textbf{CL.} & \multicolumn{1}{c|}{\textbf{OP.}} & \textbf{CL.} \\ \hline
\begin{tabular}[c]{@{}l@{}}Chain of\\ Thought\end{tabular} & \multicolumn{1}{c|}{\textbf{57 (7\%)}} & 26 (3\%) & \multicolumn{1}{c|}{\textbf{13 (2\%)}} & 7 (\textless{}1\%) & \multicolumn{1}{c|}{\textbf{43 (5\%)}} & 3 (\textless{}1\%) & \multicolumn{1}{c|}{\textbf{43 (6\%)}} & 0 \\ \hline
\begin{tabular}[c]{@{}l@{}}Directive\\ Prompting\end{tabular} & \multicolumn{1}{c|}{\textbf{68 (9\%)}} & 29 (3\%) & \multicolumn{1}{c|}{\textbf{16 (2\%)}} & 10 (1 \%) & \multicolumn{1}{c|}{\textbf{8 (1\%)}} & 5 (\textless{}1\%) & \multicolumn{1}{c|}{\textbf{2 (\textless{}1\%)}} & 1 (\textless{}1\%) \\ \hline
\begin{tabular}[c]{@{}l@{}}Responsive\\ Feedback\end{tabular} & \multicolumn{1}{c|}{\textbf{91 (12\%)}} & 16 (2\%) & \multicolumn{1}{c|}{\textbf{5 (\textless{}1\%)}} & 2 (\textless{}1\%) & \multicolumn{1}{c|}{\textbf{4 (\textless{}1\%)}} & 3 (\textless{}1\%) & \multicolumn{1}{c|}{0} & \textbf{2 (\textless{}1\%)} \\ \hline
\begin{tabular}[c]{@{}l@{}}Tree of \\ Thought\end{tabular} & \multicolumn{1}{c|}{\textbf{39 (5\%)}} & 4 (\textless{}1\%) & \multicolumn{1}{c|}{\textbf{2 (\textless{}1\%)}} & 0 & \multicolumn{1}{c|}{\textbf{10 (1\%)}} & 0 & \multicolumn{1}{c|}{0} & 0 \\ \hline
Template & \multicolumn{1}{c|}{1 (\textless{}1\%)} & 1 (\textless{}1\%) & \multicolumn{1}{c|}{0} & \textbf{1 (\textless{}1\%)} & \multicolumn{1}{c|}{0} & \textbf{1 (\textless{}1\%)} & \multicolumn{1}{c|}{0} & 0 \\ \hline
Persona & \multicolumn{1}{c|}{1 (\textless{}1\%)} & 1 (\textless{}1\%) & \multicolumn{1}{c|}{\textbf{1 (\textless{}1\%)}} & 0 & \multicolumn{1}{c|}{0} & 0 & \multicolumn{1}{c|}{0} & 0 \\ \hline
\begin{tabular}[c]{@{}l@{}}Few-shot\\ Learning\end{tabular} & \multicolumn{1}{c|}{5 (\textless{}1\%)} & \textbf{-} & \multicolumn{1}{c|}{0} & \textbf{-} & \multicolumn{1}{c|}{1 (\textless{}1\%)} & \textbf{-} & \multicolumn{1}{c|}{0} & \textbf{-} \\ \hline
\end{tabular}%
}
\vspace{-0.5cm}
\end{table}

\begin{figure}[h!]
  \centering
  \begin{subfigure}{0.4\textwidth}  
    \centering
    \includegraphics[width=\linewidth]{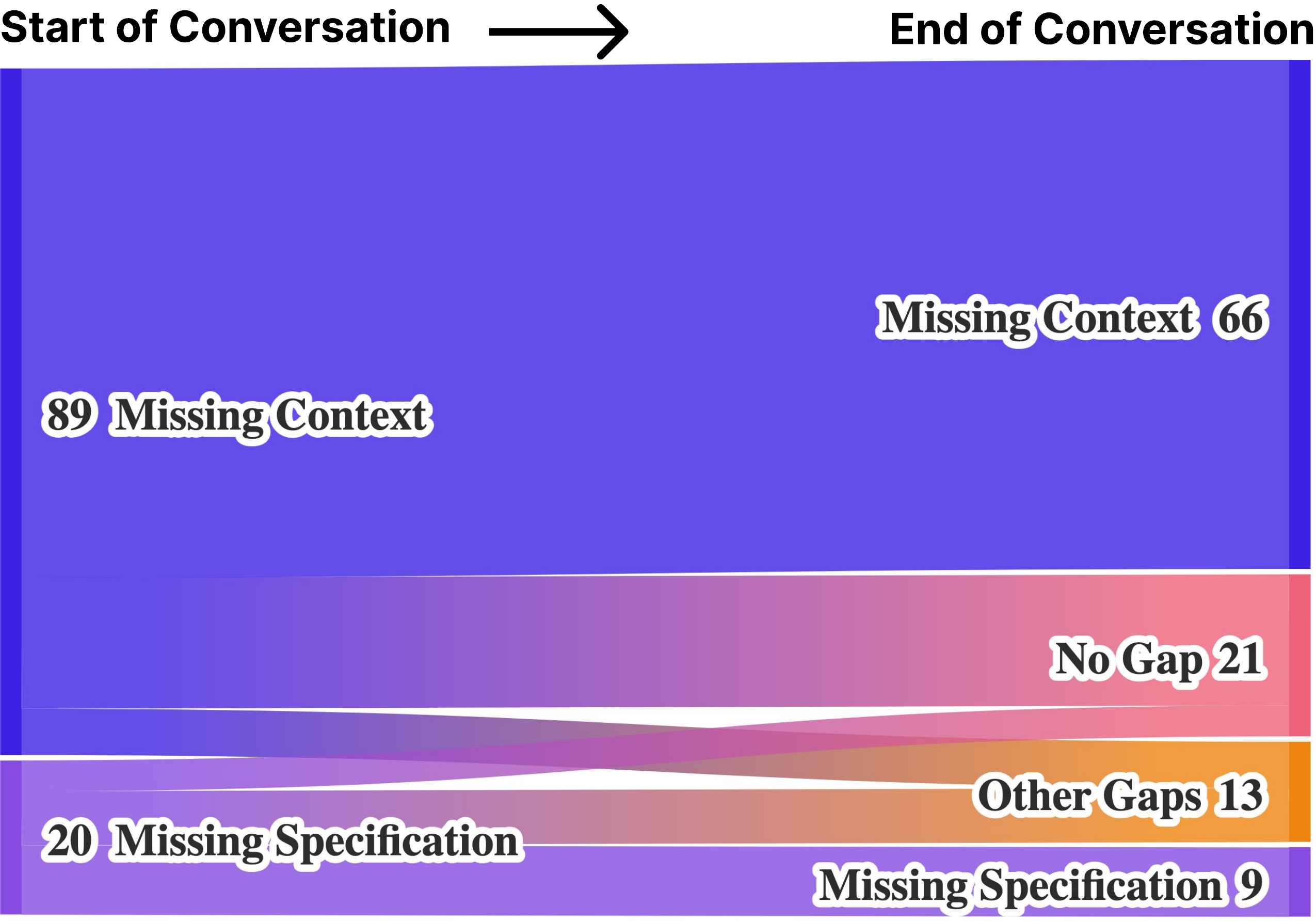}
    \caption{Open Issues}
  \end{subfigure}
  \hfill
  \begin{subfigure}{0.4\textwidth}  
    \centering
    \includegraphics[width=\linewidth]{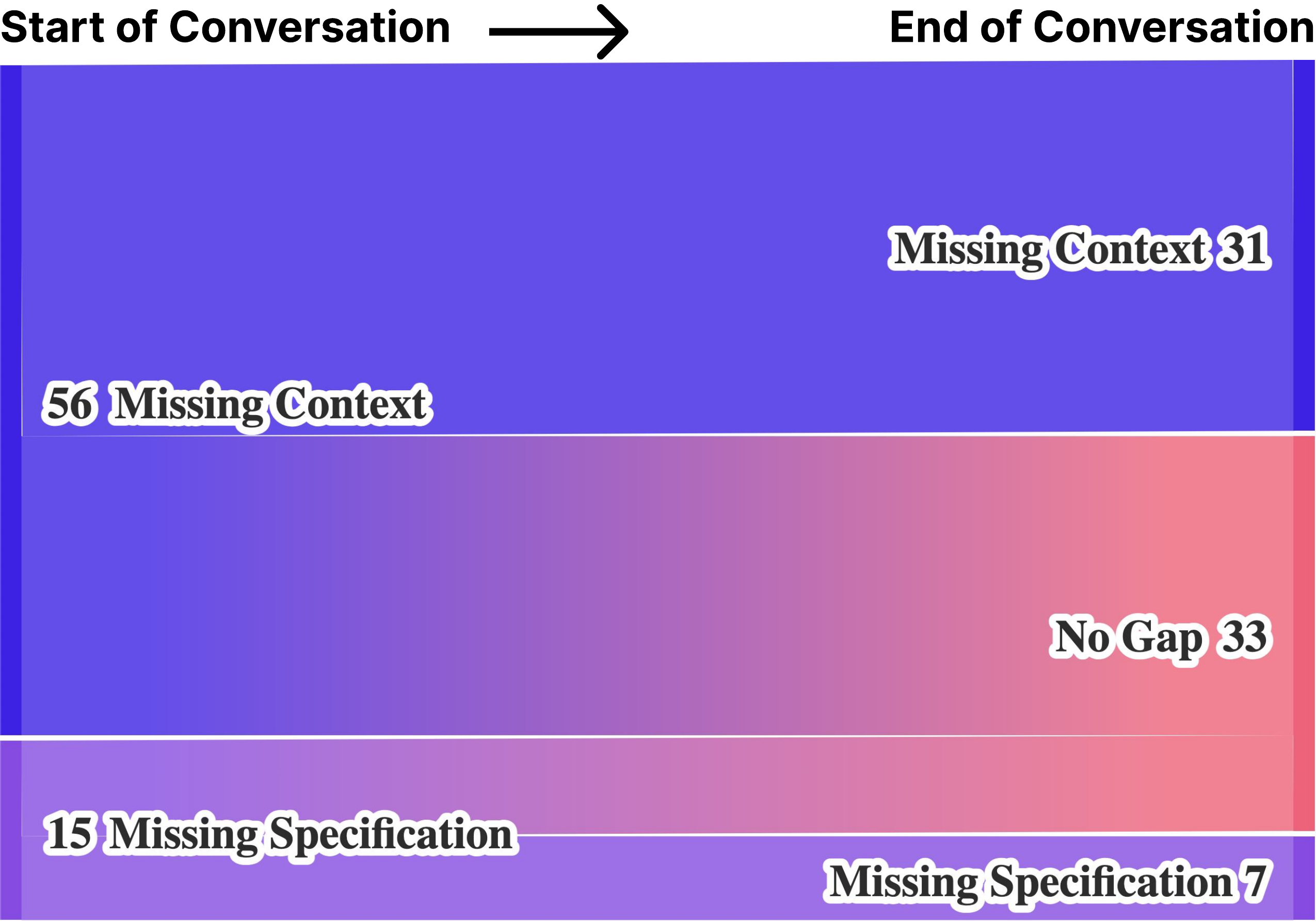}
    \caption{Closed Issues}
  \end{subfigure}
  \caption{Progression of Conversations with Prompt Knowledge Gaps}
  \label{fig:progression}
  \vspace{-0.3cm}
\end{figure}

Given that knowledge gaps are a problem in both open and closed issues—less in closed compared to open—we also looked into how developers deal with these gaps in their prompts as the conversations progress and how that determines the conversational outcome. 
We do this analysis for two of the gaps, \textit{Missing context} and \textit{Missing specification}, because these gaps can be addressed with additional rounds of interaction i.e., prompts in the same conversation. Figure \ref{fig:progression} shows the results from this analysis. 
As shown in Figure 4a, out of the 89 open conversations with \textit{Missing Context}, only 17 end with no gaps, while 72 conclude with either missing context or other gaps. 
In contrast, as shown in Figure 4b, among the 56 closed conversations that contain missing context gaps, 25 conclude with no gaps, suggesting that developers provided the necessary information and details by the end of the conversation. This shows a key difference between closed and open issues: while missing context is an issue in both, developers in closed issues tend to make more effort to provide the necessary context. 
We also observed the same pattern for \textit{Missing Specifications} in conversations. Out of 15 closed conversations that contain missing specifications (Figure 4b), 8 conversations end with no gaps. For open conversations (Figure 4a), only 4 out of 20 conversations end with no gaps. 


\noindent
\textbf{\underline{Discussion of RQ1 Findings}}

\noindent
\textbf{Style. }Across open and closed issues, developers use various conversational styles to resolve issues with ChatGPT. However, prompt knowledge gaps persist across all styles. This indicates that developers face challenges in effectively presenting issues regardless of the chosen approach. In closed issues, Directive Prompting and Chain of Thought are the two most used styles. Chain of Thought allows developers to address gaps progressively by providing additional information, but it remains highly dependent on their ability to identify and articulate missing context both timely and effectively. 

\noindent
\textbf{Gaps. }Missing Context is the most significant gap impacting issue-resolution conversations with ChatGPT. Providing accurate and relevant information is essential to help ChatGPT understand the context; failure to do so often results in misunderstandings or hallucinations. While gaps are found in both open and closed issues, closed ones tend to manage them better through additional explanations and iterative exchanges. This highlights the importance of recognizing and addressing gaps proactively when interacting with LLMs.

\subsection{RQ2: What heuristics can be used to automatically measure the prompt knowledge gaps?}
Table \ref{tab:heuristics} presents the set of textual and code-related heuristics that we investigate to automatically measure prompt knowledge gaps. 
Using Logistic Regression, we analyze the association of these heuristics with issue resolution outcomes, where closed issues indicate successful resolution. 
We chose Logistic Regression for its interpretability and efficiency in modeling binary outcomes, making it suitable for our investigation into the factors that contribute to issue resolution (i.e., open vs. closed issues).

To identify highly associated independent variables, we first calculated the Variance Inflation Factor (VIF) for our heuristics. To ensure the accuracy of our analysis, we eliminated features with a VIF greater than 5~\cite{diagnosticsLR}. The features excluded were \#Words, \#Sentences, \#Repeated 2-grams, \#Modifiers, \#SE Words, \#Distinct Words, and \#Named Entities. Additionally, to enhance the performance of our regression model, we applied an L1 penalty. Among various configurations and parameters tested, our best-performing regression model uses a Robust Scaler, an L1 penalty, and the liblinear solver with 1000 iterations. The mean Cross-Validated accuracy (CV=5) of our best-performing model is 62\%.
The coefficient values of our features are presented in Table \ref{tab:features_importance}.
In our model, the top five features with the highest coefficients are the number of misspellings in the text, the Flesch Reading Ease score, the mean size of code snippets, the number of code snippets, and the entailment of the text.  As shown in Table \ref{tab:heuristics}, we provide a range (min, median, max) for each heuristic to show their variability across conversations. To further examine the statistical significance of these heuristics, we conducted a t-test, finding several metrics to be significant (p-value\textless0.05): \#Named Entities, Mean Size of Code Snippets, \#Code Description, \#Total Prompt Count, and \#Misspellings.

\begin{figure}[h]
    \centering
    \includegraphics[width=1\columnwidth]{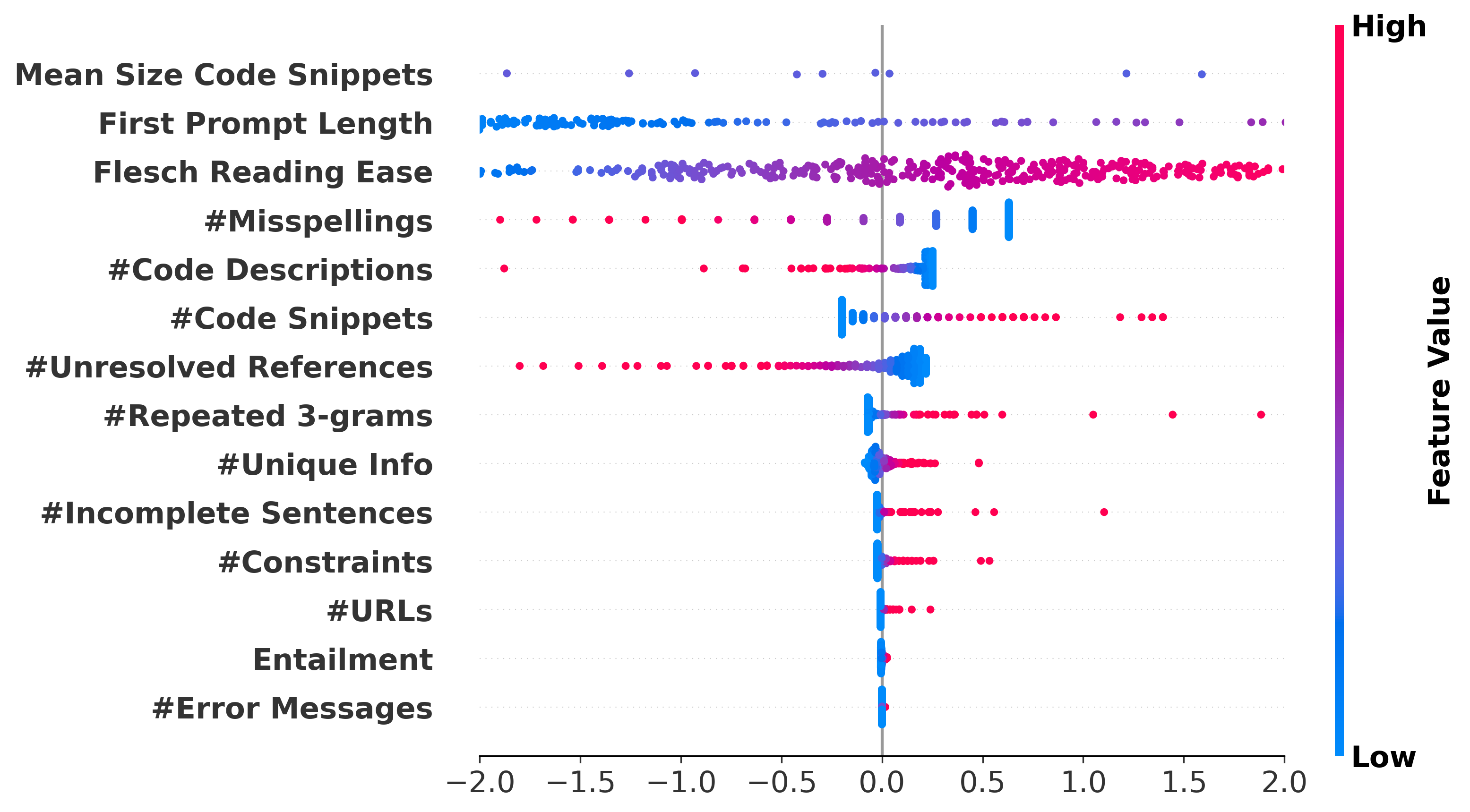}
    \caption{Impact of Features on Model's Outcome Based on SHAP}
    \label{fig:shap}
    \vspace{-0.4cm}
\end{figure}

To make our model and the effects of the features more interpretable, and signify how important each heuristic is for issue resolution, we also use SHAP to provide insights into how these features affect the model~\cite{10.5555/3295222.3295230}. The effects of each feature are shown in Figure \ref{fig:shap}. In this figure, high feature values are shown in red and low values in blue. For instance, lower values of First Prompt Length negatively impact the model (associated with open issues), while higher values positively impact it. High values of \#Unresolved References negatively affect the model.
The mean size of code snippets has a broader range than other features, making its impact less distinct in the figure due to the current x-axis limit. However, with expanded limits, high mean sizes show a strongly negative impact. Additional figures with varied x-axis limits are included in our replication package. The results of the SHAP in Figure \ref{fig:shap} and feature analysis in Table \ref{tab:features_importance} are presented below.

\noindent
\textbf{Specificity.} Effective conversations (i.e., conversations linked to closed issues) are associated with higher Information Emphasis (\#Repeated 3-grams) and Conditional Phrasing (\#Constraints). Repeated n-grams keep ChatGPT's responses aligned with the conversation's goal, maintaining the general context. Constraints, reflecting detailed specifications, help ChatGPT produce responses that are closely tailored to the developer's requests.

\noindent
\textbf{Contextual Richness.} Providing high number of code snippets (\#Code Snippets) while keeping their size small (Mean Size Code Snippets) is associated with effective issue resolution. Large code snippets can challenge ChatGPT’s limited context window, leading to less accurate responses. Including more error messages (\#Error Messages) also helps ChatGPT understand the problem context better. Additional features that are associated with effective resolution are, including references to external sources (\#URLs), unique information (\#Unique Info), and longer initial prompts (First Prompt Length). 
Even if ChatGPT cannot directly access external content, including them provides a clear indication of resources or tools relevant to the issue, which ChatGPT can factor into its response to suggest further actions.
Using repeated sets of words to maintain the context of the conversations with ChatGPT while at the same time providing more unique information to keep the conversation going forward is another interesting observation from this analysis.

\begin{table}[h!]
\caption{Coefficient Values of Features in Regression Model}
\label{tab:features_importance}
\centering
\scriptsize 
\begin{tabular}{|c|c|}
\hline
\textbf{Feature Name} & \textbf{Regression Coefficient} \\ \hline
\#Misspellings & \textbf{-0.18062388} \\ \hline
Flesch Reading Ease & \textbf{0.06966411} \\ \hline
Mean Size of Code Snippets &\textbf{ -0.06134951} \\ \hline
\#Code Snippets & \textbf{0.05318818} \\ \hline
Entailment & \textbf{0.03103776} \\ \hline
\#Unresolved Reference & -0.02926126 \\ \hline
\#Constraints & 0.02143247 \\ \hline
\#URLs & 0.01547335 \\ \hline
First Prompt Length & 0.01426245 \\ \hline
\#Code Descriptions & -0.01209851 \\ \hline
\#Incomplete Sentences & 0.0116349 \\ \hline
\#Repeated 3-grams & 0.00636866 \\ \hline
\#Unique Info & 0.00597188 \\ \hline
\#Error Messages & 0.00035634 \\ \hline
\end{tabular}%
\vspace{-0.3cm}
\end{table}

\noindent
\textbf{Clarity.} A high number of misspellings (\#Misspellings) is strongly associated with unresolved issues, highlighting their negative impact on ChatGPT conversations. High Flesch readability score (Flesch Reading Ease) and textual entailment (Entailment) further highlight the importance of clarity for effective issue resolution. Incomplete sentence structures (\#Incomplete Sentences) were not found to negatively affect conversation outcomes.

\noindent
\textbf{\underline{Discussion of RQ2 Findings}}

Our analysis highlights the importance of different heuristics that could be leveraged to automatically improve developer prompts related to issue resolution queries. Effective conversations tend to be \textit{contextually rich}, including relevant code snippets (but avoiding large files), unique details in descriptions, references, and error logs. They also exhibit \textit{high specificity} through the addition of detailed requirements related to the issue. Effective conversations also maintain \textit{clarity} by minimizing misspellings and 
ambiguity. 

%% file: sections/07_feasibility.tex
\section{Feasibility Study}
\begin{figure*}[h]
    \centering
    \includegraphics[width=0.9\textwidth]{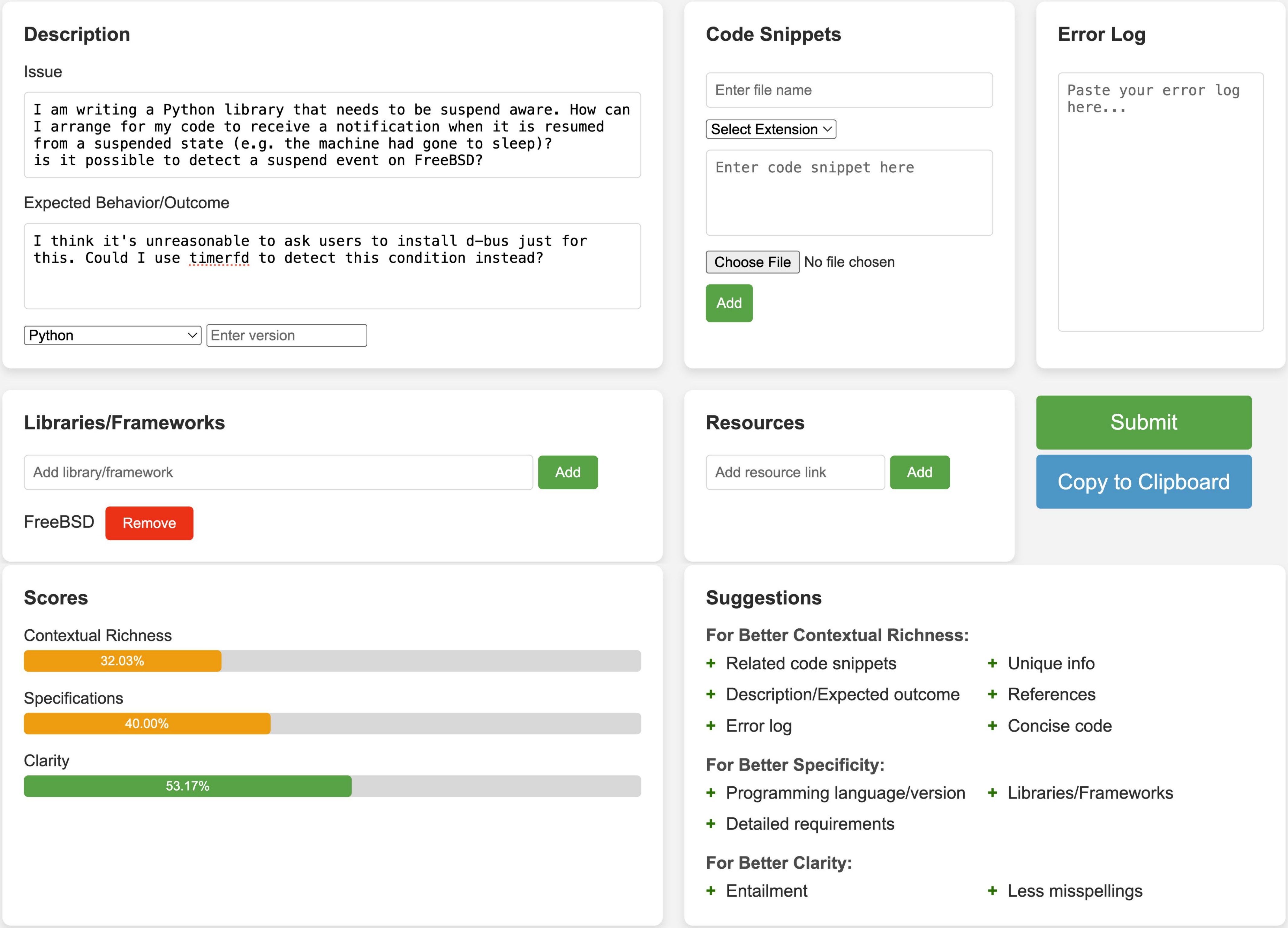}
    \caption{Tool for Automatic Prompt Knowledge Gap Detection}
    \label{fig:tool}
    \vspace{-0.4cm}
\end{figure*}

To explore whether the heuristics from RQ2 can be leveraged to develop a tool for detecting prompt knowledge gaps, we conduct a feasibility study. Our goal is to develop a lightweight tool that could be easily adapted into the developer workflows for issue resolution. 
Therefore, we develop a browser extension that can help users create tailored, detailed prompts for issue resolution.
The frontend is developed using HTML, CSS, and JavaScript, with Flask powering the backend that uses our logistic regression models to evaluate the prompts to see what is missing from the prompt. (more details are provided in the replication package). A snapshot of our tool is shown in Figure \ref{fig:tool}. Users can enter the issue details in the \textbf{Description} field, including the expected outcome, programming language, and version. Code snippets can be added by uploading files or pasting them into the \textbf{Code Snippets} box. Error logs and stack traces can be entered under \textbf{Error Log}, relevant libraries or frameworks in \textbf{Libraries/Frameworks}, and additional resources to aid context in the \textbf{Resources} box.

Our tool offers developers a structured template designed to capture critical information essential for effective issue resolution, minimizing the risk of commonly observed knowledge gaps like Missing Context, Missing Specifications, or Unclear Instructions. Once the template is completed, the tool automatically evaluates the input against predefined heuristics, generating a score for each. Based on these scores, developers can iteratively refine their inputs to achieve higher scores. Once satisfied, they can copy the optimized, structured prompt for use in their issue-resolution conversations with LLMs. To illustrate an example, we present the tool's performance using the open and closed conversations shown in Figure \ref{fig:open_vs_closed}. 
We extract the information from each conversation, populate the corresponding fields in the UI, and run the tool to display individual analyses for each case. The generated scores are the mean average of the features included in each heuristic.

\noindent
\textbf{Closed Conversation. }In this conversation, the user provides a detailed issue description, including specific requirements, a code snippet, and relevant online resources, along with libraries that the response should incorporate. After running the tool, it scores 54.07\% for contextual richness, 80\% for specificity, and 65.86\% for clarity. These scores suggest that the prompt is likely to have effective results but still has room for improvement. 
The analysis indicates that increasing the number of code snippets, and unique information inside the description can improve its context score, while improving the entailment in the description by having sentences that logically follow the same structure could enhance its clarity. Additionally, there are six misspellings that impact the clarity of the description.

\noindent 
\textbf{Open Conversation. } In this conversation (as shown in Figure \ref{fig:tool}), the user provides only a brief description of the issue, along with the programming language and framework. The tool’s analysis shows lower scores: 32.03\% for contextual richness, 40\% for specificity, and 53.17\% for clarity. The tool correctly identifies the gaps in the contextual richness of the prompt and low specifications in the requirements. Although clarity is generally good, adding relevant code snippets, resources, and a more detailed description all are required for this prompt to be more effective. These suggestions, along with the scores, are provided by the tool to guide targeted improvements.

This tool showcases the potential of leveraging heuristics to automatically identify prompt knowledge gaps in issue-resolution conversations, empowering developers to proactively enhance prompt quality. As the first step toward automating gap detection in LLM-mediated interactions, this tool lays the groundwork for advancing the quality of such conversations. However, its current capabilities are limited to predefined heuristics and may not capture all nuances of real-world developer interactions. In addition, it needs more user feedback studies to fully demonstrate its utility. Future enhancements, including additional features and more refined metrics, could further improve its effectiveness. To encourage adoption and refinement, we have made the code and usage instructions available in our replication package~\cite{anonymized}.

%% file: sections/05_threats.tex
\section{Threats to Validity}

\noindent
\textbf{Construct Validity.}
To reduce subjectivity in our annotations, we conducted multiple rounds of coding and discussions to resolve conflicts and ensure consistency. The authors performing the analysis each have over three years of experience in programming and qualitative analysis. The final average Cohen’s Kappa agreement between them was 78\%, indicating strong inter-rater reliability.

\noindent
\textbf{Internal Validity.}
The heuristics chosen for this study are based on a thorough qualitative analysis of content shared with ChatGPT for issue resolution. While these heuristics capture critical aspects, they may not encompass every nuance of the conversations. To mitigate this, we included a wide range of heuristics and removed highly correlated features using the Variance Inflation Factor (VIF) to reduce potential biases. We also conducted sanity checks on all automated measures to ensure accuracy and prevent script errors.
In our analysis, we assume that conversations in closed issues contributed to their resolution, while those in open issues did not. This assumption is reasonable, as resolved issues indicate productive exchanges. However, there might be instances where closed issues were resolved using additional help, and were not entirely based on the conversation with ChatGPT. In addition, we do not consider the difficulty of issues in our analysis. We acknowledge that further exploration of how issue complexity affects issue resolution outcomes would be valuable, and leave this as a direction for future work.

\noindent
\textbf{External Validity.}
Our findings are based on a dataset of ChatGPT conversations shared within GitHub issue threads, which may limit generalizability to other, unshared ChatGPT conversations or interactions with different LLMs, such as Gemini. We focused on ChatGPT due to its popularity and broad usage among developers. However, it is possible that a larger, more comprehensive dataset of GitHub developer-ChatGPT interactions could reveal different patterns.

%% file: sections/06_background.tex
\section{Related Work}
\label{sec:related_work}

\textbf{LLMs for Issue Resolution. }
LLMs are now widely applied for bug resolution and issue tracking \cite{hou2024large, wu2023large, tang2023empirical, 10.1145/3660773, 10.1145/3650212.3680328}. Research indicates that developers frequently engage with ChatGPT to describe bug symptoms and seek potential solutions \cite{das2024investigating}.
While Q\&A platforms like Stack Overflow have traditionally played a significant role in helping developers resolve issues, their traffic has declined with the rise of LLMs \cite{dasilva2024chatgpt}. Studies of Q\&A forum responses and LLM-generated answers reveal that LLMs struggle with inquiries related to certain frameworks and libraries \cite{dasilva2024chatgpt}. For instance, ChatGPT's accuracy for security-related questions was only 56\% \cite{10260753}. Users often prefer ChatGPT for its well-articulated language \cite{kabir2024stack}, however, ChatGPT responses are frequently of lower quality than those on Stack Overflow, often lacking relevance \cite{10298467}. In addition, answers available on Stack Overflow prove to be more effective in addressing debugging tasks \cite{liu2023better}.
Advancements in LLMs have also initiated the development of automated tools for issue resolution using LLMs~\cite{10.1109/ICSE48619.2023.00129, 10359304, 10.1145/3691620.3695537}. By compiling a dataset of GitHub issues alongside their corresponding test cases \cite{jimenez2024swebenchlanguagemodelsresolve}, researchers have evaluated various LLMs' capabilities in understanding issues and generating correct patches. For example, Tao et al. \cite{tao2024magisllmbasedmultiagentframework} introduced an LLM-powered multi-agent framework that achieved a resolution rate of 13.94\%. Subsequent studies that involved interactive sessions with LLMs to identify problematic files and relevant contextual information increased the resolution rate to 22\% \cite{10.1145/3650212.3680384}. 

\textbf{Prompt Analysis. }
Recent studies have also focused on studying interactions between software developers and ChatGPT. The most frequent inquiries directed at ChatGPT include code generation, conceptual clarifications, and how-to questions \cite{hao2024empirical}. The predominant topics identified by developers include advanced programming guidance, information-seeking about frameworks, and high-level design recommendations \cite{MohamedChatting, SagdicDiscussion}.
Developers often engage in multi-turn conversations to enhance the quality of responses by asking follow-up questions or refining their prompts \cite{hao2024empirical}. LLM-generated code are mostly used to illustrate high-level concepts or provide examples for documentation purposes. Most conversations revolve around requests for improvements and additional explanations within the generated code \cite{jin_can_2024}. Additionally, Champa et al. \cite{ChampaGHinAction} found that developers primarily seek assistance from ChatGPT for Python code related to quality management and issue resolution tasks. Developer-ChatGPT interactions have been shown to be particularly effective for software development management, optimization, and new feature implementation \cite{ChampaGHinAction}.

Mondal et al. identified 11 factors contributing to prolonged conversations with ChatGPT, with missing specifications and requests for additional functionality being the most common issues \cite{mondal2024enhancing}. While frameworks have been developed to structure prompts in various styles and techniques for improved LLM responses~\cite{white2023promptpatterncatalogenhance, ma2024saywantteaching, kim2023betterexploringpromptingstrategy}, recent studies suggest that prompt engineering is often unpredictable and unreliable, emphasizing instead the importance of clearer articulation of requests~\cite{battle2024unreasonableeffectivenesseccentricautomatic, ma2024saywantteaching}. Our paper contributes to this research by analyzing prompt knowledge gaps across conversational styles, providing heuristics to automatically detect and address these gaps to enhance issue resolution outcomes with LLMs.

%% file: sections/08_conclusion.tex
\section{Conclusion and Future Work}
LLMs have shown potential for issue resolution, but there is often a disconnect between developers’ expectations and the responses they receive. This disconnect typically arises from how issues are presented to LLMs, with insufficient context, specifications, or clarity. While frameworks and prompt-engineering methods attempt to refine LLM outputs, they remain unpredictable and largely reliant on a ``trial and error" approach. Therefore, we focus on identifying and mitigating knowledge gaps in prompts, addressing the need to help developers with targeted suggestions. 

Our analysis reveals that developers employ a range of conversational styles in issue resolution with ChatGPT, with Directive Prompting, Chain of Thought, and Responsive Feedback being the most prevalent. The most common knowledge gaps in open issues were Missing Context, Unclear Instructions, and Multiple Contexts. For closed issues, while 87.3\% of prompts contained no gaps, Missing Context remained the most frequent gap. Providing sufficient context is essential for effective resolutions, yet developers often struggle to do so effectively.
Our identified heuristics also suggest that effective conversations are contextually rich, containing related code snippets, unique information, error messages, external references, and longer initial prompts. Effective prompts also include specific requirements and technical details, as well as clear, logically structured sentences with less ambiguity.

Using these heuristics, we developed a lightweight tool to detect knowledge gaps in prompts and offer templates that guide developers in crafting contextually rich, specific, and clear prompts, enabling improved LLM-driven issue resolution. Initial design demonstrates the feasibility of such a tool, though further refinement is needed to enhance accuracy. In the future, we plan to evaluate the tool through developer feedback and questionnaires, and to experiment with additional heuristics to better capture and address prompt knowledge gaps.